\documentstyle[12pt]{article}
\begin{document}

\begin{center}

{\large \bf Neutral perfect fluids and charged thin shells\\
with electromagnetic mass in general relativity}

\vskip 1truecm

{\bf Victor Varela\footnote{Permanent Address: Centro de F\'{\i}sica
Te\'{o}rica y Computacional, Escuela de F\'{\i}sica, Facultad de
Ciencias, Universidad Central de Venezuela, Caracas,
Venezuela. E-mail: vvarela@fisica.ciens.ucv.ve}}\\

Department of Mathematical Sciences,\\
University of Aberdeen, King's College,\\
Aberdeen AB24 3UE, Scotland, UK.\\
E-mail: v.varela@maths.abdn.ac.uk\\

\end{center}

\vskip 1truecm

\abstract{We build extended sources for the Reissner-Nordstr\"{o}m
metric. Our models describe a neutral perfect fluid core bounded by
a charged thin shell, and feature everywhere positive rest mass
density and everywhere non-negative active gravitational mass, as
well as classical electron radius and electromagnetic total mass. We
contrast our results with previously discussed models featuring
similar properties at the expense of including anisotropic pressures
within the fluid. Our charged thin shells are restricted by the 2D
texture equation of state which causes the continuity of the active
gravitational mass, in spite of the singularity of the
energy-momentum tensor. We mention possible extensions of this study
suggested by modified active mass formulae proposed in the
literature.}

\vskip 1truecm

The construction of extended sources for the Reissner-Nordstr\"{o}m
(RN) geometry has interested many researchers in Einstein-Maxwell
(EM) theory. (See, for example, the partial list of contributions
reviewed by Ivanov \cite{ivanov}.) Certainly the mathematical and
physical appeal of this problem is associated with the unique
coupling of gravity to all types of matter and fields. However,
researchers often face difficulties when they use EM equations to
model the inner structure of charged particles.

Bonnor and Cooperstock  \cite{bc} used a charged fluid sphere to
model the electron, with a radius smaller than $10^{-16}$ cm. In
this study the field equations implied the existence of some
negative rest mass within the fluid (a recent overview of this
result is presented in \cite{rb}). Moreover, the total active
gravitational mass within the charged sphere was negative, which
pointed towards compatibility problems with the energy conditions
and the singularity theorems of general relativity.

Ponce de Le\'{o}n \cite{pdl1} presented a detailed analysis of
limiting configurations allowed by the energy conditions. More
recently the same author has constructed a new type of extended
source for the RN geometry, with classical electron radius
\cite{pdl2}. His relativistic version of the old
Abraham-Lorentz-Poincar\'{e} model includes a charged fluid sphere
with anisotropic pressures, a "pure field condition", and specific
continuity conditions at the surface. The appealing properties of
this electromagnetic mass model are that its active gravitational
mass is non-negative everywhere and its rest mass density is
positive everywhere. According to Ponce de Le\'{o}n, pressure
anisotropy must be considered whenever we wish to construct
electromagnetic mass sources with such properties.

The aim of this paper is to show that extended RN sources with
everywhere positive mass density, classical electron radius,
electromagnetic mass, and everywhere non-negative gravitational mass
can be constructed using perfect fluid spheres with the charge
concentrated at the surface. In our view, the use of unequal
principal pressures in \cite{pdl2} has been motivated by particular
boundary conditions, and cannot be considered as a necessary
condition to obtain the sought properties.

We shall assume a neutral perfect fluid sphere with radius $a$,
matter density $\rho$ and pressure $p$ bounded by a charged thin
shell with charge $q$, tangential stress $S$, and surface matter
density $K$. We adopt the Cohen and Cohen treatment of the EM
equations with singular energy-momentum tensor \cite{cc}, which is
suitable for the study of inner and outer spherically symmetric,
static geometries described in curvature coordinates.

Following \cite{cc} we use geometrical units ($c=G=1$) and adopt a
static, spherically symmetric line element written in the form
\begin{equation}
ds^{2}=-A^2dt^{2}+B^2dr^2+r^2\left(d\theta^2+\sin\theta^2d\phi^2\right),
\label{sch}
\end{equation}
where $A=A(r)$ and $B=B(r)$. We shall assume that $A=\sqrt{A^2}$ and
$B=\sqrt{B^2}$ are everywhere positive.

We use the orthonormal base
\begin{equation}
\begin{array}{cccc}
\hat{e}_{0}=\frac{1}{A} \frac{\partial}{\partial t},&
\hat{e}_{1}=\frac{1}{B} \frac{\partial}{\partial r},&
\hat{e}_{2}=\frac{1}{r} \frac{\partial}{\partial \theta},&
\hat{e}_{3}=\frac{1}{r \sin \theta} \frac{\partial}{\partial \phi}.
\end{array}
\label{tetrad}
\end{equation}
The field equations in this frame take the form
\begin{equation}
8\pi r^2 T^{00}=\left[r\left(1-B^{-2}\right)\right]^{\prime} ,
\label{em1}
\end{equation}
\begin{equation}
A^2 B^2 \left(1 + 8\pi r^2 T^{11}\right)=\left(rA^2\right)^{\prime},
\label{em2}
\end{equation}
\begin{equation}
8 \pi A B T^{22}=\left[\left(rB\right)^{-1}
\left(rA\right)^{\prime}\right]^{\prime} + \frac{A}{r^2 B},
\label{em3}
\end{equation}
\begin{equation}
\left(r^2 \epsilon \right)^{\prime}=4 \pi \sigma r^2 B,
\label{em4}
\end{equation}
where $T^{00}$, $T^{11}$, and $T^{22}=T^{33}$ are the non-vanishing
components of the energy-momentum tensor, prime denotes a derivative
with respect to $r$, $\sigma$ is the charge density, and $\epsilon$
is the radial electric field. We are dealing with a neutral perfect
fluid sphere bounded by a charged thin shell, so $\epsilon$ vanishes
inside the shell. Outside the shell we have ordinary vacuum
$\left(\rho = p = \sigma = 0 \right)$, and the electric field
reduces to
\begin{equation}
\epsilon = \frac{q}{r^2}, \label{coulomb}
\end{equation}
where the charge is given by
\begin{equation}
q=4\pi\int_{a_-}^{a_+}\sigma r^2 B \; dr, \label{charge}
\end{equation}
with $a_-$ and $a_+$ denoting the limit $r \rightarrow a$ taken from
below and above respectively.

The energy-momentum tensor comprises contributions from the neutral
perfect fluid, the charged thin shell and the external
electromagnetic field. It is given by
\begin{eqnarray}
T^{00}=\rho + \rho_{s} + \frac{\epsilon^2}{8\pi},& T^{11}
=p-\frac{\epsilon^2}{8\pi},& T^{22}=T^{33}=p+ p_{s}+
\frac{\epsilon^2}{8\pi},
\label{emt}
\end{eqnarray}
where $\rho$, $p$ denote the neutral fluid mass density and
pressure, respectively, and $\rho_{s}$, $p_{s}$ denote the mass
density of the shell and the tangential stress necessary to support
the shell, respectively.  In this case $\rho$ and $p$ vanish outside
the neutral fluid core, whilst $\rho_{s}$, $p_{s}$ have the forms
\begin{eqnarray}
\rho_{s}= K\; \delta\left(r-a\right),&
p_{s}=S\;\delta\left(r-a\right),
\label{deltadef}
\end{eqnarray}
where the Dirac $\delta$ function is normalized to satisfy
\begin{equation}
4\pi\int_{a_-}^{a_+}\delta\left(r-a\right) r^2 \; dr = 1.
\label{deltanorm}
\end{equation}

Cohen and Cohen use $\delta$ functions to model the thin shell
contribution to the energy-momentum tensor. Their procedure implies
formal difficulties related to the use of distributions in general
relativity. As discussed in \cite{mk}, some of these difficulties
are linked with the non-linearity of the field equations. The reader
is referred to \cite{pr} for a discussion of energy-momentum valued
distributions in the context of spherically symmetric geometries.

This thin shell formalism is characterized by its restriction to
curvature coordinates and the need to consider non-linear operations
on distributions. Nevertheless it shares with the more powerful
formalisms described in \cite{mk} the reduction of the field
equations to finite difference equations governing the
discontinuities of various quantities across the shell \cite{my}.
Cohen and Cohen's formalism provides us with formulae for the
determination of the physical parameters $K$ and $S$ as well.

Now we combine Eqs. (\ref{em1})-(\ref{em3}) with Eq. (\ref{emt}) and
integrate the arising equations in different spacetime regions. In
the outer region we impose the appropriate weak field behavior for
large $r$ to obtain the RN solution
\begin{eqnarray}
A^{2}=B^{-2}=1-\frac{2m}{r}+\frac{q^2}{r^2},& r>a, \label{RN}
\end{eqnarray}
where $m$ is the total mass of the solution. We add the restriction
$a>2m$ to guarantee that $A^2$ and $B^2$ are positive for $r>a$.

From Eqs. (\ref{em1})-(\ref{em3}) we get
\begin{equation}
4\pi r^2 A B \left(T^{00}+T^{11}+T^{22}+T^{33}\right) =
\left(\frac{r^2 A^{\prime}}{B}\right)^{\prime}.
\label{eqfun}
\end{equation}

Restricting our attention to the inner region, and substituting Eq.
(\ref{emt}) into Eq. (\ref{eqfun}) we obtain
\begin{equation}
4\pi r^2 A B \left(\rho + 3p\right) = \left(\frac{r^2
A^{\prime}}{B}\right)^{\prime}, \label{eqfun1}
\end{equation}
which suggests a particular choice for the equation of state of the
neutral perfect fluid.

The significance of the equation of state
\begin{equation}
\rho+3p=0 \label{eosn}
\end{equation}
was discussed by several authors in connection with global monopoles
and textures (see \cite{d} and references therein). In a recent
paper \cite{v}, the simplifying power of this equation of state has
been highlighted in the context of neutral perfect fluid,
Majumdar-type solutions of EM equations. In this paper we deal with
neutral perfect fluids as well and assume that Eq. (\ref{eosn}) is
valid throughout the core. If $B(0)$ is finite and
$A^{\prime}(0)=0$, then we integrate Eq. (\ref{eqfun1}) to get
\begin{eqnarray}
A=A_{0}=constant,& r<a. \label{resA}
\end{eqnarray}

Combining Eqs. (\ref{em1}), (\ref{em3}) and (\ref{emt}) with Eqs.
(\ref{eosn}) and (\ref{resA}), we find that $h=\frac{1}{r^2 B^2}$
satisfies $h^{\prime}=-\frac{2}{r^3}$ and obtain
\begin{eqnarray}
B^2=\frac{1}{1-C r^2},& r<a, \label{resB}
\end{eqnarray}
where $C$ is an integration constant. Using the same field equations
combined with Eqs. (\ref{resA}) and (\ref{resB}) we get
\begin{eqnarray}
\rho=\frac{3C}{8\pi},& p=-\frac{C}{8\pi}, \label{denspre}
\end{eqnarray}
for the constant mass density and pressure of our neutral perfect
fluid core.

The internal solution given by Eqs. (\ref{resA}) and (\ref{resB})
has vanishing Weyl tensor and constant Ricci scalar. When
$0<C<\frac{1}{a^2}$ it actually describes a section of a spherical
Einstein universe, and $B_{-}^2=\left(1-Ca^2\right)^{-1}$ is bounded
from above.

Integration of Eq. (\ref{em1}) across the shell yields
\begin{equation}
\frac{1}{B_{-}^{2}}-\frac{1}{B_{+}^2}=\frac{2K}{a}, \label{Bjump}
\end{equation}
where $B_{+}^2=\left(1-\frac{2m}{a}+\frac{q^2}{a^2}\right)^{-1}$.
Thus $B^2$ is discontinuous at $r=a$ whenever $K$ is non zero, and
Eq. (\ref{Bjump}) implies the identity
\begin{equation}
m=\frac{Ca^3}{2}+K+\frac{q^2}{2a}, \label{massformula}
\end{equation}
where the provisionally independent contributions of the fluid core,
the shell, and the electromagnetic field to the total mass are
specified.

Since both $B_-^2$ and $B_+^2$ are bounded from above, Eq.
(\ref{em2}) implies the continuity of $A^2$ at $r=a$. Thus we
combine Eqs. (\ref{RN}) and (\ref{resA}) to get
\begin{equation}
A_0^2=1-\frac{2m}{a}+\frac{q^2}{a^2}.
\end{equation}

Integration of Eq. (\ref{em3}) across the shell yields
\begin{equation}
8\pi S \int_{a_-}^{a_+} B \delta\left(r-a\right) dr =
\frac{1}{B_+}\left(\frac{1}{a}+\frac{A_+^\prime}{A_0}\right)
-\frac{1}{B_-}\left(\frac{1}{a}+\frac{A_-^\prime}{A_0}\right).
\label{intS}
\end{equation}
Cohen and Cohen pointed out that although the integral on the left
of Eq. (\ref{intS}) is not well defined, it can be indirectly
determined by integrating Eq. (\ref{em1}) across the shell. This
procedure yields the exact result
\begin{equation}
\int_{a_-}^{a_+} B \delta\left(r-a\right) dr = \frac{1}{4\pi a K}
\left(\frac{1}{B_-}-\frac{1}{B_+}\right), \label{intBd}
\end{equation}
which, combined with Eq. (\ref{intS}), provides us with a formula
for the elastic stress $S$. The straightforward manipulation of an
ill defined distributional operation -via the field equations- is
indeed a striking feature of Cohen and Cohen's thin shell formalism.

Zloschchastiev \cite{z1} has determined equilibrium conditions for
spherical charged thin shells enclosing ordinary vacuum. Using
Israel's standard formalism \cite{i}, he considers shells with
surface mass densities and pressures satisfying linear equations of
state. Motivated by Zloschchastiev's work on 2D textures
\cite{z1,z2}, we impose the equation
\begin{equation}
K+2S=0 \label{ees}
\end{equation}
and study its implications for a charged shell surrounding a neutral
perfect fluid core. (We point out that Cohen and Cohen's paper
\cite{cc} does not make any reference to possible equations of state
satisfied by the shell parameters.)

Combining Eqs. (\ref{intS}) and (\ref{intBd}) we derive a clumsy
equation for $S$ as a function of $a$, $q$, $C$, $K$ that can be
simultaneously solved with Eq. (\ref{ees}). This procedure provides
us with an expression for the surface matter density, namely,
\begin{equation}
K=-\frac{1}{2}Ca^3+\frac{q^2}{2a}. \label{ecuaK}
\end{equation}
Substitution into Eq. (\ref{massformula}) yields
\begin{equation}
m=\frac{q^2}{a}. \label{classical}
\end{equation}
Thus our charged sources have classical electron radius. Notably Eq.
(\ref{classical}) is assumed from the outset in the very different
approaches developed in \cite{pdl2} and \cite{z1}.

Using Eqs. (\ref{denspre}) and (\ref{ecuaK}) we see that the
restriction $0 < C < \frac{q^2}{a^4}$ guarantees the existence of a
family of solutions with $\rho > 0$ and $K>0$. Additionally, the
charged shell that encloses ordinary vacuum (case $C=0$) is
characterized by $K=\frac{q^2}{2a}>0$.

If $a^2 > q^2$ then $0 < C < \frac{q^2}{a^4} \Rightarrow 0 < C <
\frac{1}{a^2}$ and we are dealing with useful solutions, i.e. the
ones with $\rho>0$, $K>0$, and bounded $B_-$. We point out that the
restrictions $a>2m$ and $a^2>q^2$ are compatible with the
approximate model of the electron discussed in \cite{bc}. (However,
it should be noted that the corresponding classical radius
determined with Eq. (\ref{classical}) is about $2817$ times larger
than the experimental upper limit of $10^{-16}$ cm.)

Ponce de Le\'{o}n \cite{pdl2} uses the Tolman-Whittaker formula
\begin{equation}
M(r)=4\pi \int_{0}^{r}(T^{00}+T^{11}+T^{22}+T^{33})\, A\, B\, r^2\,
dr \label{TW}
\end{equation}
to determine the active gravitational mass inside a sphere of radius
$r$. Gr{\o}n \cite{g} justified the use of the surface gravity
$\kappa$, given by
\begin{equation}
\kappa=-\frac{M(r)}{r}, \label{sg}
\end{equation}
as a measure of (coordinate) radial acceleration for geodesic motion
at instantaneous rest. The later author derived Eqs. (\ref{TW}) and
(\ref{sg}) from the geodesics and Einstein equations, and discussed
them when a singular shell is present. In this case a difference
between the values of $M(r)$ at both sides of the shell is expected,
and the integration is performed through the shell. Gr{\o}n also
pointed out that negative values for $M(r)$ cause repulsive
gravitation effects that may play a role in connection with
elementary particle models. (See \cite{bgms} for another recent
consideration of Eq. (\ref{TW}) in connection with RN spacetimes.)

We combine the Tolman-Whittaker formula with the Cohen and Cohen
method in order to determine the active gravitational mass of our
solutions, including the contribution of the charged thin shell. In
fact, taking into account the useful solutions discussed above and
Eq. (\ref{TW}), we obtain
\begin{eqnarray}
M(r)=0,& r<a,
\end{eqnarray}
and
\begin{eqnarray}
M(r)=q^2 \left( \frac{1}{a}-\frac{1}{r} \right),& r>a.
\end{eqnarray}

We observe that neither the neutral perfect fluid core nor the
charged thin shell have active gravitational mass as a consequence
of equations of state (\ref{eosn}) and (\ref{ees}). Thus $M(r)$ is
continuous at $r=a$, and the Maxwell field contributes with positive
active gravitational mass for $r>a$. Additionally, $M(r)$ has the
limit $M(\infty)=m$.

The extended sources for the RN metric discussed here include a
neutral perfect fluid core bounded by a charged thin shell. The mass
density $\rho$ and the surface charge density $K$ are strictly
positive, and the gravitational mass $M(r)$ is everywhere
non-negative. The discontinuity imposed by the thin shell prevents
the consideration of anisotropic pressures within the fluid core.
Repulsive gravitational forces are totally irrelevant to these
solutions.

In the limit $q \rightarrow 0$ the physical parameters $\rho$, $p$,
$K$ and $S$ as well as $m$ vanish. Thus, the total gravitational
mass of our solutions is built from electromagnetism only. Other
electromagnetic mass sources discussed in the literature
\cite{trk,gau} contain negative active mass and exert repulsive
gravitational forces on test particles \cite{g}.

Using Eq. (21) in \cite{pdl2}, we see that the combination of Eq.
(\ref{eosn}) with the conformal flatness of the internal metric
determines the vanishing of the active gravitational mass for $r<a$.
This interior geometry makes (coordinate) acceleration as well as
tidal forces for radial geodesic motion vanish \cite{d}. On the
other hand, the 2D charged texture equation of state (\ref{ees})
suffices for the vanishing of the thin shell gravitational mass and
the consequent continuity of $M(r)$ at $r=a$. (Whether the automatic
vanishing of the 3D Weyl tensor plays a role in the vanishing of the
active gravitational mass of the shell with equation of state
(\ref{ees}) is a point that deserves further attention.) Considering
that the total energy-momentum tensor is singular at $r=a$, the
continuity of $M(r)$ is an intriguing feature of these models.

The above discussion is based on Eq. (\ref{TW}). Reservations about
the use of this formula have been expressed in the past, due to its
lack of invariance under a change of scale of the time coordinate
\cite{bonnor}. On the other hand, this formula may lead to wrong
results when discontinuity surfaces are included in the solutions.
These observations motivated Devitt and Florides \cite{DeFlor} to
develop modified mass formulae that correct the time invariance
problem and deal with static, spherically symmetric thin shells in
accordance with Israel's standard formalism. A study of the useful
solutions based on the Tolman-Whittaker-Devitt-Florides formalism
should clarify the nature of these extended sources for the RN
metric.

\section*{Acknowledgements}
The author is grateful to the University of Aberdeen for hospitality
while this paper was being written, and to Professor G.S. Hall for a
number of useful suggestions and comments. He also thanks Mr. M.
Chung for computational assistance. This work was partially
supported by grants from the Universidad Central de Venezuela (CDCH
and Vicerrectorado Acad\'{e}mico), the Royal Society of Edinburgh,
and the Edinburgh Mathematical Society.

\end{document}